\documentclass{article}

\usepackage{microtype}
\usepackage{graphicx}
\usepackage{subcaption}
\usepackage{subcaption}
\usepackage{booktabs} 
\usepackage{amsmath}
\usepackage{xspace}

\usepackage{hyperref}

\newcommand{\tool}{Lumos\xspace}

\usepackage[accepted]{mlsys2025}

\mlsystitlerunning{\tool: Efficient Performance Modeling and Estimation for Large-scale LLM Training}

\begin{document}

\twocolumn[
\mlsystitle{\tool: Efficient Performance Modeling and Estimation for Large-scale LLM Training}



\mlsyssetsymbol{equal}{*}

\begin{mlsysauthorlist}
\mlsysauthor{Mingyu Liang}{cornell}
\mlsysauthor{Hiwot Tadese Kassa}{meta}
\mlsysauthor{Wenyin Fu}{meta}
\mlsysauthor{Brian Coutinho}{meta}
\mlsysauthor{Louis Feng}{meta}
\mlsysauthor{Christina Delimitrou}{mit}
\end{mlsysauthorlist}

\mlsysaffiliation{cornell}{Cornell University, Ithaca, USA}
\mlsysaffiliation{meta}{Meta, Menlo Park, USA}
\mlsysaffiliation{mit}{Massachusetts Institute of Technology, Cambridge, USA}

\mlsyscorrespondingauthor{Mingyu Liang}{ml2585@cornell.edu}

\mlsyskeywords{Large Language Models, Performance Modeling, Distributed Training, Simulation}

\vskip 0.3in

\begin{abstract}

Training LLMs in distributed environments presents significant challenges due to the complexity of model execution, deployment systems, and the vast space of configurable strategies. Although various optimization techniques exist, achieving high efficiency in practice remains difficult. Accurate performance models that effectively characterize and predict a model’s behavior are essential for guiding optimization efforts and system-level studies. We propose \tool, a trace-driven performance modeling and estimation toolkit for large-scale LLM training, designed to accurately capture and predict the execution behaviors of modern LLMs. We evaluate \tool on a production ML cluster with up to 512 NVIDIA H100 GPUs using various GPT-3 variants, demonstrating that it can replay execution time with an average error of just 3.3\%, along with other runtime details, across different models and configurations. Additionally, we validate its ability to estimate performance for new setups from existing traces, facilitating efficient exploration of model and deployment configurations.

\end{abstract}
]



\printAffiliationsAndNotice{}  

\section{Introduction}

In recent years, large language models (LLMs) have transformed many aspects of daily life. The availability of vast datasets, along with advancements in computational resources, has enabled the development of increasingly complex models, such as ChatGPT~\cite{ouyang2022training}, LLaMA~\cite{touvron2023llama}, and PaLM~\cite{chowdhery2023palm}. However, efficiently training these LLMs presents significant challenges, necessitating both hardware and software innovations across the system stack.

To meet these demands, efforts have focused on addressing various bottlenecks. Key areas of optimization include the development of AI-specific hardware (e.g., NVIDIA GPUs~\cite{NVIDIA_blackwell} and SmartNICs~\cite{ma2022fpga}), improvements in memory systems~\cite{kwon2023efficient, rajbhandari2020zero}, the design of optimal parallelism strategies~\cite{zheng2022alpa, isaev2023calculon}, overlapping communication with computation~\cite{hashemi2019tictac, narayanan2021efficient}, and advancements in algorithms~\cite{beltagy2020longformer, you2019large}.

Despite these innovations, ensuring training efficiency remains a significant challenge. Diagnosing inefficiencies in LLMs is particularly difficult because runtime traces produced by machine learning (ML) frameworks~\cite{PyTorch, TensorFlow} are often dense and require deep expertise to interpret effectively. Moreover, runtime behavior can vary significantly across different model architectures, deployment configurations, accelerator types, network infrastructures, and other system components. These variations can cause performance bottlenecks to shift unpredictably, making them difficult to identify and address. An additional challenge lies in the vast search space of optimization possibilities. Finding the optimal solution within this space is time-consuming and resource-intensive, as it requires extensive experimentation on real hardware, demanding significant resources and incurring high costs.

A key step toward achieving efficiency is to accurately characterize and understand the behavior of these models. One common approach is to build performance models that capture model execution, which also provide a solid foundation for further optimization studies. While existing efforts~\cite{moolchandani2023amped, isaev2023calculon} develop analytical models to predict high-level performance based on exposed model parameters, they often miss essential underlying execution details. To address this limitation, recent work~\cite{hu2022dpro, zhu2020daydream, lin2022building, bang2023vtrain} has leveraged runtime traces to construct fine-grained execution graphs, providing deeper insights into the execution process.

\begin{figure}[htb]
    \centering
    \includegraphics[width=0.48\textwidth]{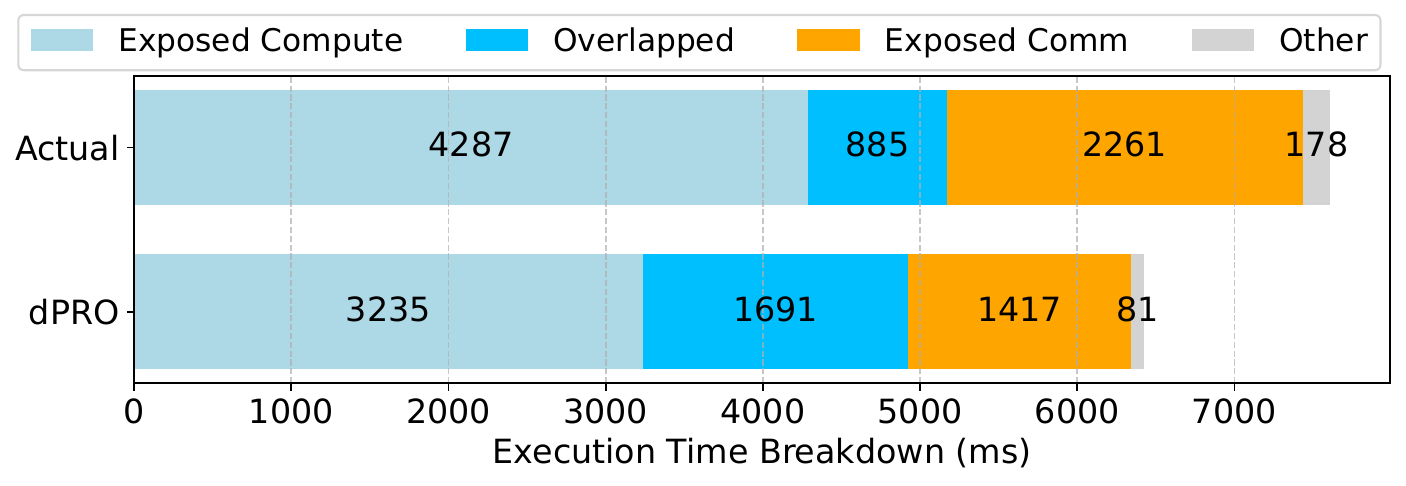}
    \caption{Execution breakdown for one training iteration of GPT-3 175B, configured with tensor parallelism = 8, pipeline parallelism = 4, and data parallelism = 8.}
    \label{fig:execution_breakdown}
\end{figure}

However, current modeling methods struggle to address the new complexities in modern LLMs. Training LLMs at scale involves deploying across multiple machines, introducing substantial communication overhead. As one example of optimization, overlapping computation with communication can reduce end-to-end training time, but it also introduces complex inter-stream dependencies that are challenging to model accurately. Figure~\ref{fig:execution_breakdown} shows the execution time breakdown for a single training iteration of the GPT-3 (175B) model, along with replayed results from dPRO~\cite{hu2022dpro}, a state-of-the-art trace-driven performance modeling tool. The comparison reveals significant gaps between the simulated and actual results, highlighting the challenges in capturing the full intricacies of LLM execution.

To overcome these difficulties, we propose \tool, a trace-driven performance modeling and estimation toolkit for large-scale LLM training. To the best of our knowledge, \tool is the first system to provide accurate performance models that effectively capture the execution behaviors of LLMs. It leverages built-in profiling tools from ML frameworks, such as PyTorch Kineto~\cite{pytorch-kineto}, without requiring any custom instrumentation in models or frameworks, thereby minimizing the profiling overhead.

Furthermore, to streamline the exploration of optimization opportunities, \tool also offers the flexibility to modify and generate new execution graphs from existing traces. This capability facilitates the exploration of optimal configurations, such as adjusting parallelism strategies (e.g., data and pipeline parallelism) and fine-tuning model architectures (e.g., number of layers, hidden size). By estimating performance through simulation rather than experimenting on real hardware, \tool can significantly reduce cost and accelerate the optimization process.

The main contributions of our work are the following:

\begin{itemize}
\item By utilizing only built-in profiling traces from ML frameworks, \tool constructs a comprehensive execution graph that identifies all dependencies between executed tasks, enabling accurate performance modeling of large-scale LLM training. Beyond estimating overall execution time, the fine granularity of \tool allows it to reproduce detailed execution characteristics, facilitating deeper analysis and downstream optimization studies.

\item With a detailed execution graph, \tool offers users a convenient way to explore various model configurations, including adjustments to parallelism strategies and model architectures. By manipulating the existing graph to generate new ones for different configurations and by predicting performance through simulation, \tool streamlines the optimization process and enables efficient and low-cost configuration exploration. 

\item We evaluate \tool using various GPT-3 model variants on a production ML cluster with up to 512 NVIDIA H100 GPUs. Our results show that \tool can accurately replay execution, achieving an average error of only 3.3\% across different models and deployment configurations. Additionally, we demonstrate that \tool accurately reproduces detailed execution statistics, such as execution time breakdown and SM utilization, showing significant improvements over existing approaches. Finally, we validate its ability to estimate performance for new configurations and deployments, achieving high accuracy when adjusting parallelism strategies and tuning various model architectures.
\end{itemize}

\section{Related Work}

\subsection{Profiling Tools and Traces}

As the ML system stack evolves rapidly, profiling tools play a crucial role in understanding model execution characteristics and identifying performance bottlenecks. As hardware accelerators like GPUs~\cite{NVIDIA_blackwell} and TPUs~\cite{jouppi2023tpu} become increasingly essential, vendors offer specialized tools—such as NVProf~\cite{NVProf}, CUPTI~\cite{CUPTI}, and Nsight~\cite{Nsight}—to expose hardware performance counters, providing developers with critical insights into performance metrics and enabling effective optimization.

To improve the interpretability of profiling results, ML frameworks also provide built-in tools for collecting execution statistics at the operator level. These tools often integrate hardware-level traces, offering a complete view of the entire stack—from host to device. For instance, PyTorch Kineto~\cite{pytorch-kineto} leverages CUPTI~\cite{CUPTI} to capture runtime information for PyTorch operators, CUDA events, and GPU kernels, seamlessly linking them to provide a holistic perspective on model execution.

\subsection{LLMs and Parallelism Strategies}

Most modern LLMs are built on transformer architectures~\cite{vaswani2017attention}, which rely on self-attention mechanisms to capture long-range dependencies in sequential data. These models feature multiple stacked layers of attention and feedforward networks, with parameter sizes growing rapidly over the years. For example, GPT-2~\cite{radford2019language} introduced in 2019 had 1.5 billion parameters, GPT-3~\cite{brown2020language} in 2020 expanded to 175 billion parameters, and PaLM~\cite{chowdhery2023palm} reached 540 billion parameters by 2022.

Training LLMs presents significant computational and memory challenges, especially as model sizes grow beyond the capacity of individual GPUs. To address these limitations, 3D parallelism—a hybrid approach combining data, tensor, and pipeline parallelism—has become essential for efficient large-scale training~\cite{narayanan2021efficient, shoeybi2019megatron, smith2022using, chowdhery2023palm}. Each form of parallelism contributes uniquely: data parallelism (DP) distributes training batches across devices, synchronizing gradients during updates; tensor parallelism (TP) splits large tensors across multiple GPUs, allowing shared computation with frequent communication; and pipeline parallelism (PP) partitions the model into sequential stages, with each stage processed on different devices in a coordinated pipeline.

Despite the benefits, configuring 3D parallelism introduces significant complexity, requiring careful coordination across these strategies to balance workloads and minimize communication overhead. Recent research has focused on automating these configurations to reduce the burden on developers and ensure efficient distributed execution. For example, GSPMD~\cite{xu2021gspmd} extends the XLA compiler~\cite{sabne2020xla} to support various parallelism paradigms through user annotations. Alpa~\cite{zheng2022alpa} automates model parallelization by optimizing intra- and inter-operator parallelism for efficient distributed execution. Galvatron~\cite{miao2022galvatron} introduces a decision tree to decompose the search space and designs a dynamic programming algorithm to generate the optimal plan.

Emerging techniques like sequence parallelism~\cite{li2021sequence, jacobs2023deepspeed, liu2023ring} further address the challenges of training on long sequences by distributing computations along the sequence dimension, reducing memory overhead and communication bottlenecks.

\subsection{Performance Modeling, Simulation, and Optimization}

The complexity of LLMs poses challenges and opportunities in system design and optimization, with performance modeling serving as a critical foundation for diagnosing and optimizing overall efficiency.

There are two primary approaches to building performance models. The first relies on analytical models. AmPeD~\cite{moolchandani2023amped} introduces an analytical model to estimate performance in distributed transformer training under various model parameters and parallelism strategies. Similarly, Calculon~\cite{isaev2023calculon} provides a parameterized analytical model that explores the co-design space of software and hardware configurations to identify optimal system designs for LLMs. However, these analytical models are often tailored to specific implementations and hardware configurations, limiting their ability to generalize in the face of rapid model and system evolution. Moreover, they typically provide high-level performance estimates, making them inadequate for optimizations like mixed precision training~\cite{das2018mixed, zhu2020daydream} and operator fusion~\cite{zhao2022apollo, jia2019taso}.

The second approach leverages trace-based models to simulate execution and derive optimization insights. For example, ASTRA-sim~\cite{rashidi2020astra} and ASTRA-sim2.0~\cite{won2023astra} simulate distributed training with a cycle-level and analytical network backend, evaluating collective communication algorithms and network topologies. In ~\cite{lin2022building}, the authors analyze critical paths within profiled traces to predict per-batch training time for DLRM. Daydream~\cite{zhu2020daydream} uses kernel-level dependency graphs collected with CUPTI to predict runtime under specific optimizations, while dPRO~\cite{hu2022dpro} builds a global dataflow graph by tracking dependencies among operators to estimate DNN training performance. However, these trace-based approaches fail to fully capture the complexities inherent in LLM execution. To the best of our knowledge, this work is the first to leverage traces for accurately modeling the intricate behaviors of LLMs, accounting for detailed operator and kernel interactions essential for precise performance prediction.

\section{Design}

\subsection{Overview}

\begin{figure*}[htb]
    \centering
    \vspace{0.1in}
    \includegraphics[width=\textwidth]{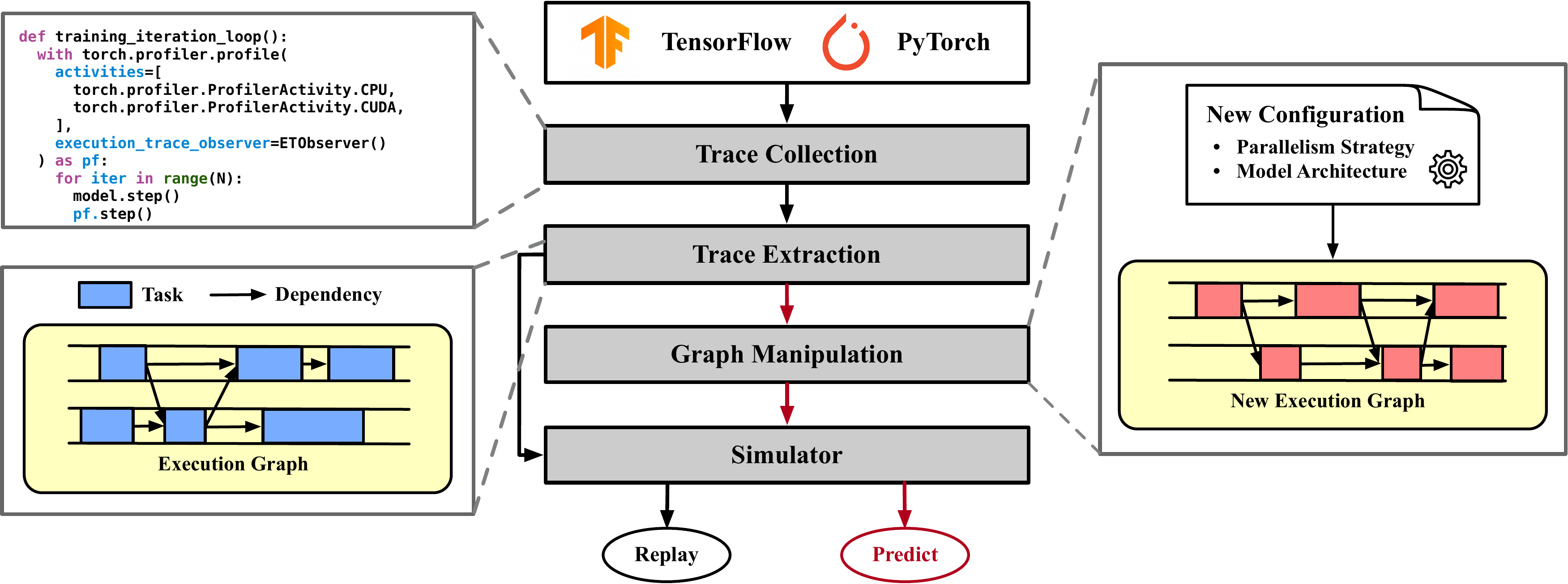}
    \caption{Overview of \tool's workflow. }
    \label{fig:overview}
\end{figure*}

Figure~\ref{fig:overview} presents the workflow of \tool, our trace-driven performance modeling and estimation toolkit for distributed LLM training. The process begins with collecting runtime profiling traces from popular frameworks such as TensorFlow and PyTorch. These raw traces are then analyzed to extract key meta-information, which is used to construct a detailed task-level execution graph. The execution graph can be modified to adjust model configurations, such as parallelism strategies and architectures, generating new configuration-specific graphs. Finally, the simulator uses these graphs to either replay the original execution or predict performance under alternative configurations, providing insights into potential optimizations and enabling effective exploration of what-if scenarios.

We initially focus on PyTorch due to its widespread use in both academia and industry, along with its advanced profiling capabilities. However, our approach is flexible by design and can be extended to support other ML frameworks. We will discuss the adaptability of \tool in Section~\ref{sec:discussion}.

\subsection{Traces Collection}

We collect profiling traces using PyTorch Kineto~\cite{pytorch-kineto}, which captures comprehensive runtime information about PyTorch operators, CUDA runtime events, and GPU kernels, including name, start time, duration, CUDA stream ID, thread ID, correlation ID, and more. Unlike previous methods such as Daydream and dPRO, which necessitate extensive framework and model instrumentation, our profiling involves adding only a few lines of code into the model, as shown in the code snippet at the top left of Figure~\ref{fig:overview}, significantly improving usability with minimal effort.

\subsection{Execution Graph}
\label{sec:execution graph}

The essence of a model’s execution lies in its execution graph, which maps out the tasks being performed and the dependencies between them. Motivated by prior approaches~\cite{zhu2020daydream, hu2022dpro, bang2023vtrain}, we construct a low-level execution graph to accurately represent model execution. However, we have incorporated several enhancements to capture the complex execution characteristics of LLMs, ensuring more accurate modeling and offering the flexibility to estimate performance for new model configurations.

\subsubsection{Tasks}

To streamline the design, our execution graph includes only the following two types of tasks:

\textbf{CPU tasks}: These include all executions happened on the CPU, including PyTorch operators and CUDA runtime events. For each CPU task, we record its metadata along with the specific CPU thread on which it runs.

\textbf{GPU tasks}: These include all executions happened on the GPU, which primarily consist of GPU kernels. For each GPU task, we log its metadata along with the corresponding CUDA stream responsible for its execution.

\subsubsection{Dependency}
\label{sec:task_dependency}

\begin{figure}[htb]
    \centering
    \includegraphics[width=0.48\textwidth]{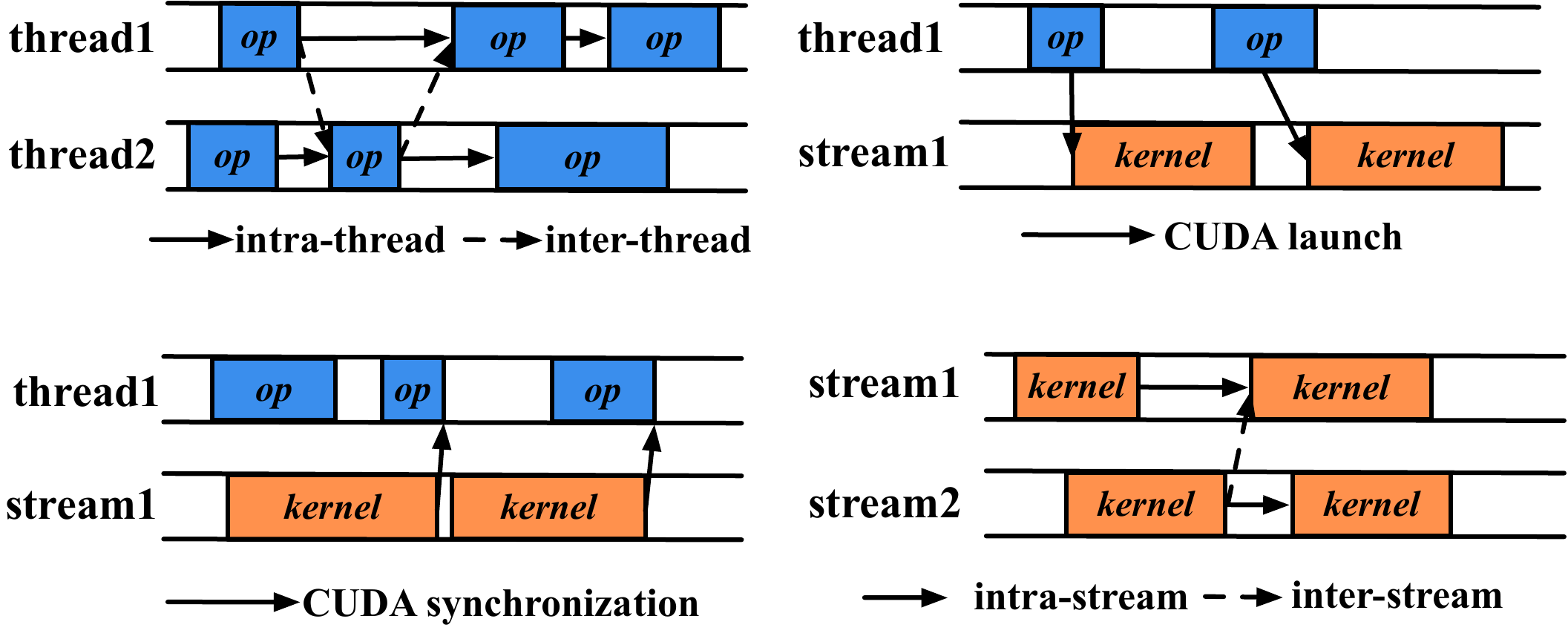}
     \caption{Four types of dependencies between the tasks.}
    \label{fig:dependency}
\end{figure}

Next, we identify four types of dependencies that capture all possible relationships between tasks:

\textbf{CPU to CPU}: This dependency includes both intra-thread and inter-thread relationships between CPU tasks. Tasks within the same thread naturally execute sequentially, forming intra-thread dependencies between consecutive tasks. Inter-thread dependencies occur when tasks on one thread block tasks on another. For example, in PyTorch, the backward pass runs on a separate thread, requiring the first backward operator to wait until the last forward operator completes. We detect these dependencies by identifying significant execution gaps within threads and establishing cross-thread dependencies accordingly.

\textbf{CPU to GPU}: GPU tasks are typically launched by corresponding CPU-side CUDA events, such as \texttt{cudaLaunchKernel} and \texttt{cudaMemsetAsync}. In Kineto traces, both CUDA runtime events and GPU kernels are tagged with a correlation ID, which we use to link CPU tasks with their corresponding GPU tasks.

\textbf{GPU to CPU}: CUDA synchronization events, such as \texttt{cudaDeviceSync} or \texttt{cudaStreamSync}, are common during model execution. When invoked on the CPU, these events block execution until the relevant GPU kernels complete. As a result, they create dependencies from one or more GPU tasks to the initiating CPU task.

\textbf{GPU to GPU}: Similar to CPU-to-CPU dependencies, this includes both intra-stream and inter-stream dependencies between GPU tasks. GPU kernels within the same CUDA stream execute sequentially, meaning consecutive tasks in the same stream have direct dependencies. To identify inter-stream dependencies, we leverage a specialized event-based synchronization mechanism captured in the Kineto trace. Specifically, we focus on a pair of CUDA runtime events: \texttt{cudaEventRecord} and \texttt{cudaStreamWaitEvent}. The \texttt{cudaEventRecord} marks a synchronization point in one stream, recording an event after all preceding kernels on that stream have completed. The corresponding \texttt{cudaStreamWaitEvent} ensures that a different stream waits until the recorded event is reached, creating a dependency between the two streams. This mechanism allows us to accurately capture inter-stream dependencies, providing a precise representation of the execution order across streams.

Training LLMs at scale typically spans a large number of machines, resulting in significant communication overhead. To mitigate this, overlapping the execution of computation and communication kernels is a common practice to reduce end-to-end iteration time. However, this overlap introduces complex inter-stream dependencies, which are overlooked by existing modeling approaches. \tool is the first to target LLMs and capture their intricate dependencies, a critical step toward accurate performance modeling and reliable downstream optimization studies.

\subsection{Graph Manipulation}

To improve and optimize LLM training performance, researchers and engineers can have many configurable options and optimization strategies. Commonly, they will ask what-if questions, such as: 

\textit{How will the performance scale with additional GPUs? Which parallelism configuration will deliver the best results? How will changes to the model architecture impact performance? Will a specific optimization improve performance, and by how much?}

While current distributed training frameworks make it easier to change configurations, deploying models with new settings on real hardware requires substantial resources and incurs high costs, leading to long iteration cycles. The process becomes even more challenging if the desired changes, such as introducing a new operator fusion pattern, are not supported by the framework, forcing developers to hack the underlying code, which can be both time-consuming and prone to errors.

To address these challenges, the fine granularity and flexibility of execution graphs, combined with simulation, provide an effective solution. By modifying the existing graph to reflect different model configurations, we can estimate performance and explore what-if scenarios without requiring large-scale physical deployments, accelerating iteration speed and significantly reducing costs. 

\tool offers an interface that allows users to specify new model configurations, after which it manipulates the existing execution graph to generate a new one reflecting the changes for performance estimation. It currently supports modifications to both model architectures—such as adjusting the number of transformer layers and hidden size—and parallelism strategies, including data parallelism and pipeline parallelism.

\begin{figure}[h]
    \centering
    \includegraphics[width=0.50\textwidth]{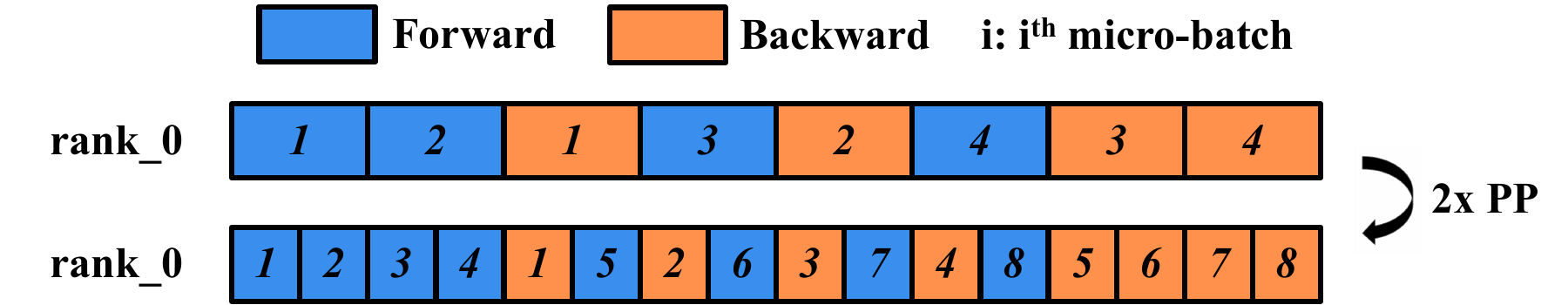}
    \caption{Updated pipeline schedule for rank\_0 with 2x PP, assuming the number of micro-batches is equal to TP × PP and 1F1B scheduling policy~\cite{narayanan2021efficient}. }
    \label{fig:pipeline_schedule}
\end{figure}

For changes in data parallelism, only the communication needs adjustment by assigning new execution time to the communication tasks, as the local computation for each worker remain unchanged. For pipeline parallelism adjustments, we first update the pipeline schedule to align with the new configuration based on the scheduling policy, determining the execution order of the forward and backward passes, as illustrated in Figure~\ref{fig:pipeline_schedule}. Next, we group the tasks by layers and partition the original layers and their underlying tasks into new stages. For example, assuming layers are evenly distributed, we calculate how many layers belong to each stage. The corresponding tasks are reassigned to their new stages, and communication tasks are inserted at appropriate points, to ensure correct synchronization and execution.  We currently do not support modifications to tensor parallelism, as it is typically fixed in practice (e.g., within a single node) due to its high communication overhead. We leave the support for it as our future work.

For changes to model architecture, such as adjusting the hidden size, we modify the input tensor dimensions for the relevant operators and kernels and update their execution times during simulation. When changing the number of layers, we follow a process similar to that used for pipeline parallelism, dividing tasks into layers and applying the adjustments accordingly.

Throughout this manipulation process, we ensure that the dependency patterns from the original trace are preserved in the new graph to maintain correct execution, as we will demonstrate in the evaluation.

\subsection{Simulation}

\begin{algorithm}[htb]
\caption{\tool's Simulation Algorithm}
\label{alg:sim}
\begin{algorithmic}
\STATE \textbf{Input:} Execution graph: G = (V, E)
\STATE \textbf{Output:} Trace with runtime details of all tasks

\STATE $\mathcal{R} \gets \emptyset$  \COMMENT{Initialize the ready task set}
\STATE $P \gets \{0\}$  \COMMENT{Initialize task processors}

\FOR{each task $t \in G.tasks$}
    \STATE $t.dep \gets |\{t's \text{ fixed dependencies}\}|$
    \IF{$t.dep = 0$}
        \STATE $\mathcal{R} \gets \mathcal{R} \cup \{t\}$
    \ENDIF
\ENDFOR

\WHILE{$\mathcal{R} \neq \emptyset$}
    \STATE $t \gets pick(\mathcal{R})$  \COMMENT{Select a ready task to execute}
    \STATE $p \gets t.Processor$
    \STATE $\mathcal{R} \gets \mathcal{R} \setminus \{t\}$

    \FOR{each $r \in get\_runtime\_dependencies(t)$}
        \STATE $r.dependents \gets r.dependents \cup \{t\}$
        \STATE $t.dep \gets t.dep + 1$
    \ENDFOR

    \IF{$t.dep = 0$}
        \STATE $t.start \gets \max(P[p], t.start)$
        \STATE $P[p] \gets t.start + t.duration$

        \FOR{each $c \in t.dependents$}
            \STATE $c.dep \gets c.dep - 1$
            \STATE $c.start \gets \max(c.start, t.start + t.duration)$
            \IF{$c.dep = 0$}
                \STATE $\mathcal{R} \gets \mathcal{R} \cup \{c\}$
            \ENDIF
        \ENDFOR 
    \ENDIF
\ENDWHILE
\end{algorithmic}
\end{algorithm}

Both the original and modified execution graphs will be fed into the simulator to simulate execution and estimate performance. During the simulation, the four types of dependencies outlined in Section~\ref{sec:task_dependency} are maintained through two mechanisms. Fixed dependencies are determined at initialization and remain unchanged throughout execution, such as the sequential order of CPU tasks on the same thread. Runtime dependencies, on the other hand, are assigned dynamically during runtime. For example, a \texttt{cudaStreamSync} task must wait for the last kernel on a specific stream to complete, but which kernel will be last cannot be known prior to execution.

Algorithm~\ref{alg:sim} outlines the simulation process, beginning with the assignment of fixed dependencies. In each iteration, a ready task is selected and allocated to its respective processor. The algorithm then checks for any runtime dependencies. If all dependencies are met, the task is executed, updating the processor’s progress and the status of its dependent tasks. Otherwise, the task is deferred until all dependencies are resolved. The simulation generates a trace similar to the input trace initially profiled from the real run, recording all runtime information of the tasks. This output trace can be used not only to estimate the overall execution time but also to analyze fine-grained execution characteristics, as we show in Section~\ref{sec:replay}.

\section{Evaluation}

We implement \tool in Python with approximately 5,200 LoC. To leverage it, users need access to the source code to insert profiler hooks into their PyTorch models for collecting traces, typically requiring around 10 lines of code. \tool then offers a fully automated workflow: it begins by constructing the execution graph from the raw traces, manipulates the graph based on new configurations, and concludes with performance estimation through simulation. Depending on the complexity of the original traces, the entire process can range from a few seconds to several minutes.

\subsection{Methodology}

\textbf{Models.} We evaluate \tool using NVIDIA's open-source GPT-3 implementation~\cite{GPT-3} from the MLPerf Training Benchmarks. Our experiments involve training four model variants, adjusting the number of layers, hidden size, feedforward network size, and attention heads, with model parameters ranging from 15 billion to the full 175 billion, as summarized in Table~\ref{tab:model_arch}. We collect traces with PyTorch Kineto and evaluate performance across various parallelism strategies, exploring different combinations of tensor, pipeline, and data parallelism.

\textbf{Infrastructure.} Our evaluation is conducted on a production ML cluster, using up to 512 NVIDIA H100 GPUs (on 32 servers) interconnected with 8x 400Gbps per host in a RoCE DC-scale network. Our testing environment is based on CUDA 12.4, PyTorch 2.5, Transformer Engine 0.12.0, and PyTorch Lightning 1.9.4. 

We select dPRO~\cite{hu2022dpro} as the state-of-the-art baseline for comparison. In our evaluation, we first validate the replay accuracy by comparing both iteration time and execution breakdown against the ground truth and the baseline. Next, we evaluate the accuracy of our approach in estimating performance for new configurations, including changes in both parallelism strategies and model architectures.

\begin{table*}[t]
\caption{Model sizes and architectures used in the evaluation. All other parameters follow the default values from the open-source GPT-3 implementation~\cite{GPT-3}.}
\label{tab:model_arch}
\vskip 0.15in
\begin{center}
\begin{sc}
\begin{tabular}{lcccccr}
\toprule
Model Name & $n_{params}$ & $n_{layers}$ & $d_{model}$ & $d_{ffn}$ & $n_{heads}$ & $d_{head}$\\
\midrule
GPT-3 15B     & 15B & 48 & 6144 & 12288 & 48 & 128  \\
GPT-3 44B     & 44B & 48 & 12288 & 24576 & 48 & 128  \\
GPT-3 117B    & 117B & 96 & 12288 & 24576 & 96 & 128  \\
GPT-3 175B    & 175B & 96 & 12288 & 49152 & 96 & 128  \\
\bottomrule
\end{tabular}
\end{sc}
\end{center}
\vskip -0.1in
\end{table*}

\begin{figure*}[h]
    \centering
    \includegraphics[width=0.99\textwidth]{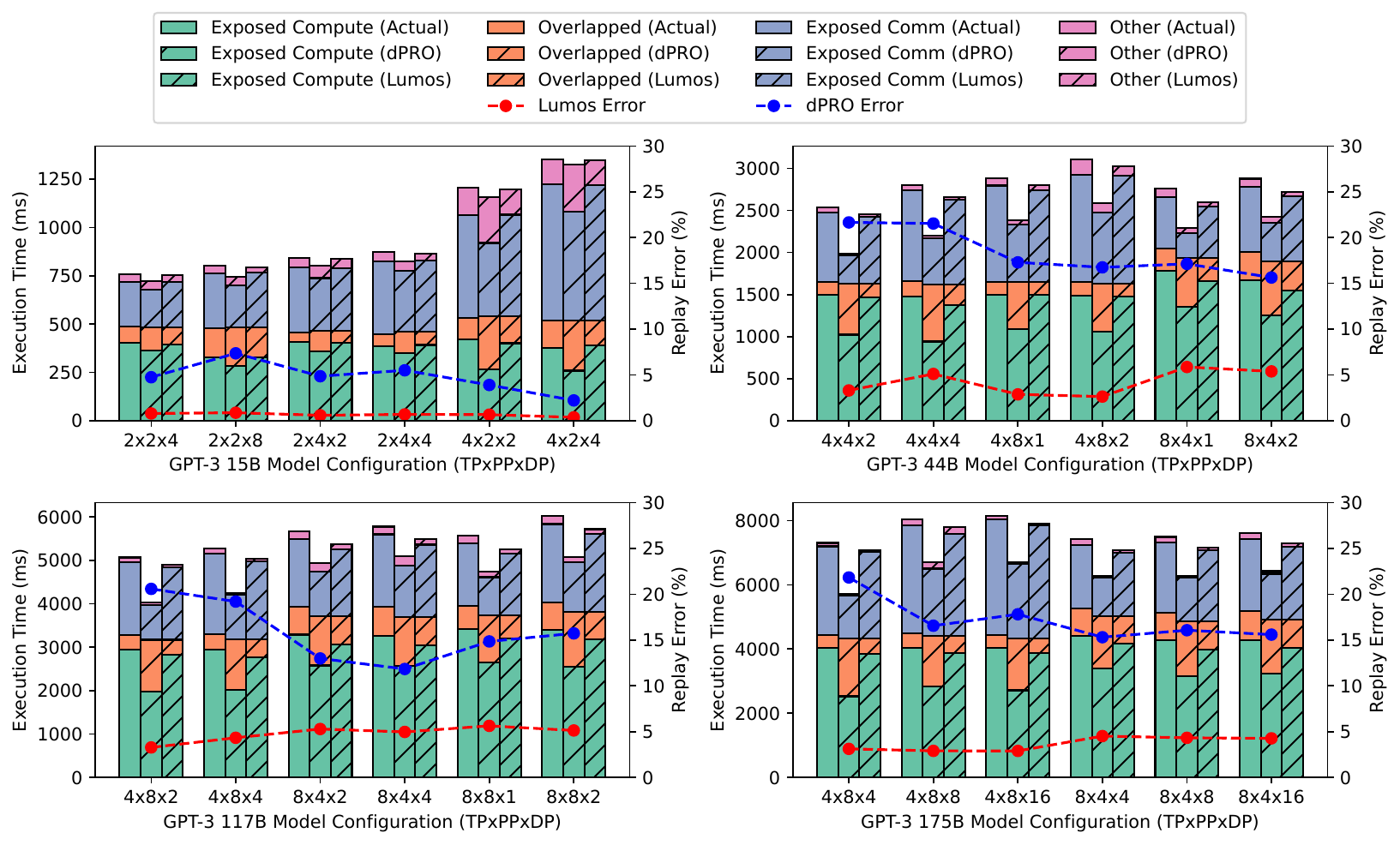}
    \caption{Per-iteration training time with its breakdown across various model sizes and parallelism strategies: comparison of actual execution, dPRO, and \tool. }
    \label{fig:replay}
\end{figure*}

\subsection{Replay}
\label{sec:replay}

\subsubsection{Overall Iteration Time}
\label{sec:overall_iteration_time}

In Figure~\ref{fig:replay}, we firstly compare the per-iteration execution time replayed by \tool and dPRO against the real execution time measured from actual training across various GPT-3 model sizes (15B, 44B, 117B, and 175B) and parallelism strategies. Across all configurations, \tool maintains a replay error mostly under 5\%, with an average error of 3.3\%. In contrast, dPRO’s error reaches up to 21.8\%, with an average of 14\%. While smaller models and simpler setups allow dPRO to predict the overall time relatively well, its accuracy deteriorates as model size and complexity grow. The discrepancy highlights \tool's robustness in accurately capturing execution behaviors and modeling performance, even for larger models and complex deployment setups.

\subsubsection{Execution Breakdown}
\label{sec:execution_breakdown}

Figure~\ref{fig:replay} also provides a detailed breakdown of the execution into key components: exposed compute (computation that does not overlap with communication), exposed communication (communication that does not overlap with computation), overlapped execution (where computation and communication run concurrently), and other (primarily idle periods). This breakdown offers deeper insights into the differences in iteration times across configurations. 

The analysis reveals that dPRO consistently overestimates overlapped execution and underestimates total iteration time, primarily due to its inability to accurately model inter-stream dependencies, leading to overly optimistic predictions of parallel execution. In contrast, \tool effectively captures the complex dependencies within the model and faithfully replays the execution. It accurately reflects the dynamic interactions between computation and communication, adapting to changes in model size and deployment configuration, and closely aligning with the real measurement.

In large-scale distributed training, particularly for LLMs, a substantial portion of execution time is spent on communication and synchronization between GPUs. To optimize performance, engineers aim to maximize the overlap between computation and communication kernels. Therefore, an accurate performance model that not only replays overall execution time but also captures fine-grained details, such as the degree of overlap, is essential. Such a model can provide valuable insights for identifying performance bottlenecks and guiding optimization efforts.

\subsubsection{SM Utilization}

\begin{figure}[h]
    \centering
    \includegraphics[width=0.48\textwidth]{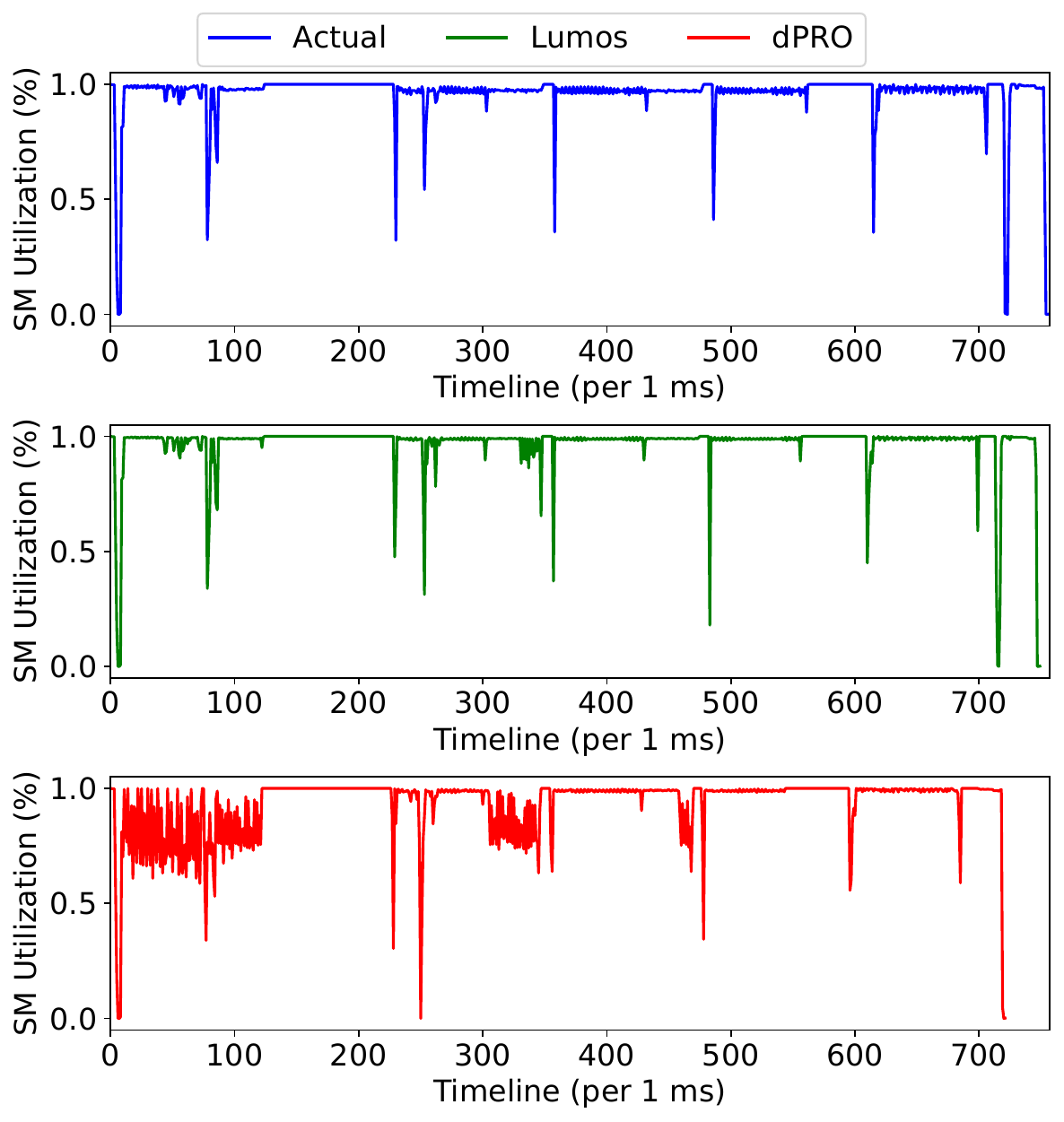}
    \caption{SM utilization of one iteration when training GPT-3 15B under TP = 2, PP = 2 and DP = 4.}
    \label{fig:sm_utilization}
\end{figure}

Analyzing SM (Streaming Multiprocessor) utilization is essential for identifying performance bottlenecks, such as idle periods or imbalanced workloads, to enhance GPU efficiency. 

In this section, we examine the SM utilization over one iteration of training the GPT-3 15B model, configured with tensor parallelism = 2, pipeline parallelism = 2, and data parallelism = 4. Utilization is defined as the fraction of time, over 1ms intervals, during which at least one CUDA stream is actively executing tasks. This data is derived from profiled and simulated traces by analyzing kernel activities throughout the execution.

As shown in Figure~\ref{fig:sm_utilization}, the SM utilization replayed by \tool closely match the actual measured utilization. In contrast, dPRO exhibits more fluctuations and significant discrepancies. This comparison, again, highlights \tool's ability to capture fine-grained execution details, validating its effectiveness in accurately modeling execution behavior.

\subsection{Graph Manipulation}

\subsubsection{Parallelism Strategy}
\label{sec:parallel}

\begin{figure}[htb]
\centering
\begin{subfigure}[b]{0.48\textwidth}
    \centering
    \includegraphics[width=\textwidth]{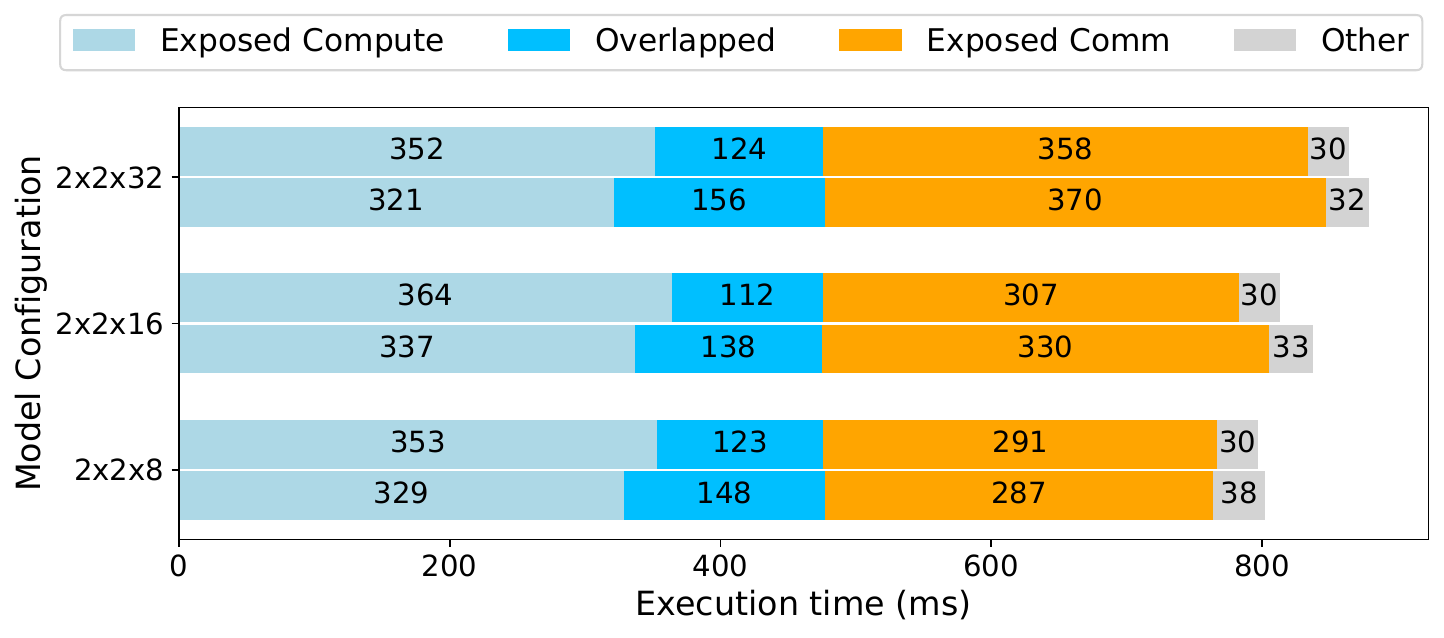}
    \caption{Execution breakdown for scaling data parallelism. }
    \label{fig:scale_dp}
\end{subfigure}
\hfill
\begin{subfigure}[b]{0.48\textwidth}
    \centering
    \includegraphics[width=\textwidth]{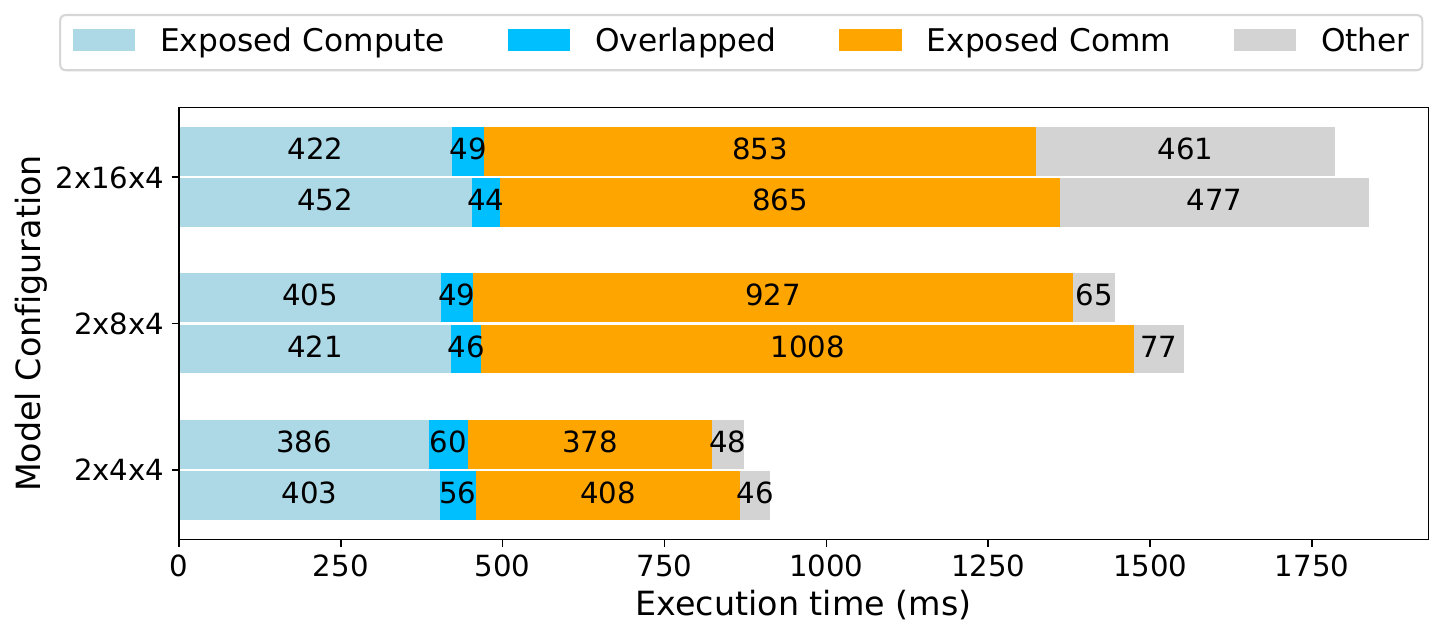}
    \caption{Execution breakdown for scaling pipeline parallelism.}
    \label{fig:scale_pp}
\end{subfigure}
\hfill
\begin{subfigure}[b]{0.48\textwidth}
    \centering
    \includegraphics[width=\textwidth]{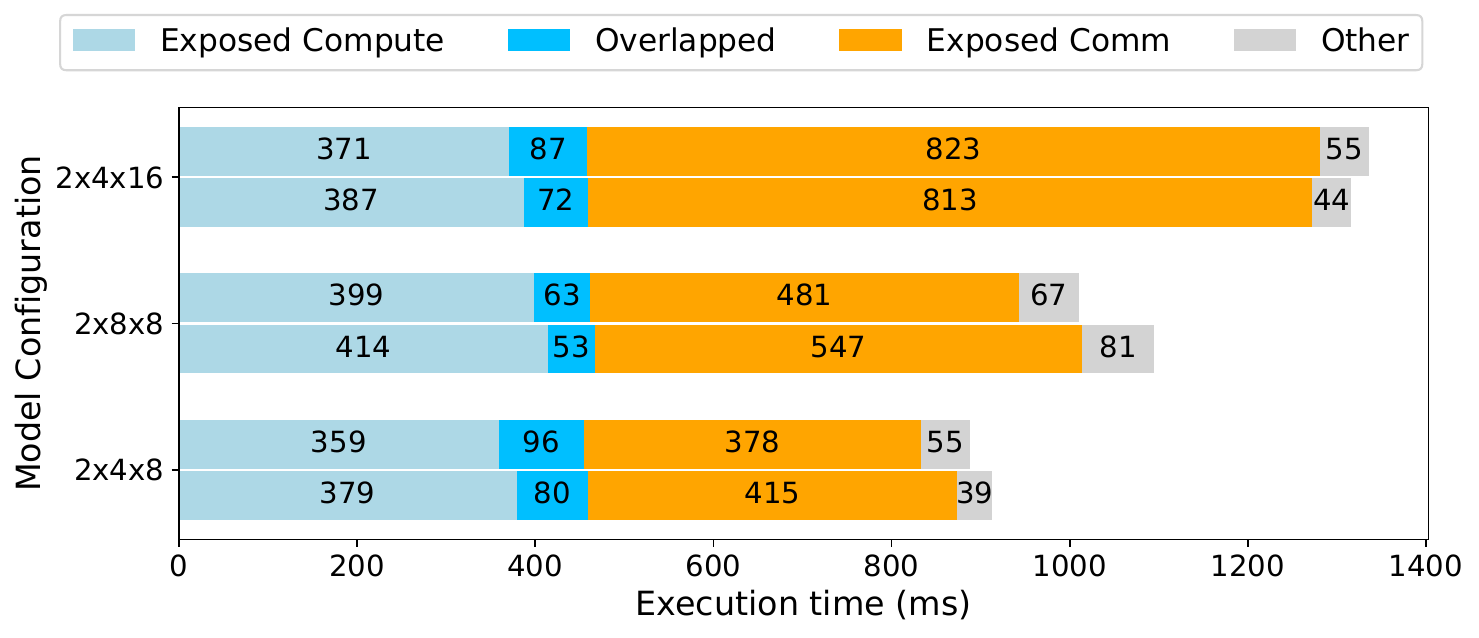}
    \caption{Execution breakdown for simultaneous scaling of data and pipeline parallelism.}
    \label{fig:scale_together}
\end{subfigure}
\caption{Runtime predictions for scale-out configurations. Each configuration (TPxPPxDP) is represented by two horizontal bars: the upper bar shows the predicted value by \tool, and the lower bar shows the actual value. }
\label{fig:scale_up}
\end{figure}

Next, we evaluate \tool's flexibility in generating new execution graphs from existing ones to estimate performance under new configurations. Specifically, we demonstrate its ability to predict scale-out performance by adjusting parallelism strategies. Our experiments focus on GPT-3 15B, using traces collected from a baseline configuration with tensor parallelism = 2, pipeline parallelism = 2, and data parallelism = 4.

We begin by exploring changes to data parallelism, where only the execution time of communication tasks needs to be updated. We currently estimate new communication time using an in-house performance model built from fleet traces, as it is both readily available and accurate. While network simulators like ASTRA-sim~\cite{won2023astra} or analytical models~\cite{moolchandani2023amped, rashidi2022themis} could also be used, predicting the runtime of individual kernels is beyond the scope of this work. We will explore potential integrations with these tools in the discussion section. To validate these predictions, we compare them against actual traces collected at larger scales. As shown in Figure~\ref{fig:scale_dp}
, \tool accurately predicts both the total runtime and detailed performance breakdowns when scaling from 16 GPUs to 32, 64, and 128 GPUs. 

Similarly, Figure~\ref{fig:scale_pp} demonstrates that \tool can also accurately estimate execution time and breakdown when scaling pipeline parallelism. We modify the baseline traces by splitting layers and underlying tasks into new stages, adding communication tasks, and reordering task execution according to the new pipeline schedule. Finally, Figure~\ref{fig:scale_together} shows that \tool maintains high accuracy, with an average error of just 4.2\% when scaling both data and pipeline parallelism simultaneously. These results prove that \tool can effectively generate correct new executions through graph manipulation for new parallelism strategies.


\subsubsection{Model Architecture}
\label{sec:model_arch}

\begin{table}[h]
\caption{Sizes and architectures for model variations.}
\label{tab:model_variation}
\vskip 0.15in
\begin{center}
\begin{small}
\begin{sc}
\begin{tabular}{lcccr}
\toprule
Model Name & $n_{params}$ & $n_{layers}$ & $d_{model}$ & $d_{ffn}$ \\
\midrule
GPT-3 15B       & 15B & 48 & 6144 & 12288    \\
GPT-3 v1       & 20B & 64 & 6144 & 12288   \\
GPT-3 v2       & 30B & 96 & 6144 & 12288   \\
GPT-3 v3       & 28B & 48 & 9216 & 18432   \\
GPT-3 v4       & 44B & 48 & 12288 & 24576   \\
\bottomrule
\end{tabular}
\end{sc}
\end{small}
\end{center}
\end{table}

We now validate \tool's accuracy in estimating performance for different model architectures. Our evaluation continues with GPT-3 15B as the base model. To generate several variants, we modify the number of layers, hidden sizes, and feedforward network sizes. Table~\ref{tab:model_variation} summarizes the sizes and architectures of the models used in this evaluation. All models are trained using the configuration of tensor parallelism = 2, pipeline parallelism = 2, and data parallelism = 4.

When increasing the number of layers, we duplicate the layers and corresponding tasks from the existing trace, insert them into the graph at appropriate places, and reconstruct dependencies with neighboring tasks according to the original dependency pattern. For changes in hidden size or feedforward network size, we adjust the input tensor dimensions for all relevant operators and kernels. Ideally, new execution times should be assigned to all affected tasks to reflect the input changes. However, we observe that only a few key kernels, such as GEMM and communication-related ones, exhibit significant runtime changes under different configurations. We similarly update the execution times for these kernels using the in-house performance model described in Section~\ref{sec:parallel}.

\begin{figure}[htb]
    \centering
    \includegraphics[width=0.482\textwidth]{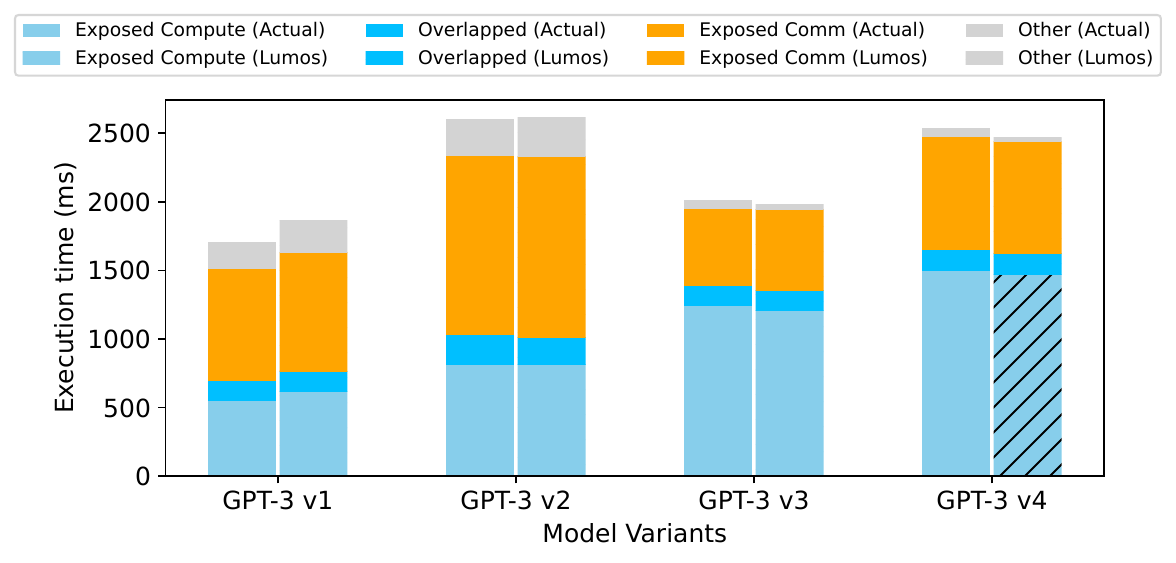}
    \caption{Iteration time breakdown of model variations. The left bars show the actual values, and the right hatched bars show the predicted values.}
    \label{fig:arch_breakdown}
\end{figure}

Figure~\ref{fig:arch_breakdown} presents the iteration time breakdown across these model variations, showing both the actual and predicted performance. The predicted values, represented with hatched patterns, align closely with the actual measurements, demonstrating that \tool accurately reproduces the execution and estimates the performance under different model architecture changes. 

Overall, the results shown in Section~\ref{sec:parallel} and Section~\ref{sec:model_arch} demonstrate \tool's ability to leverage existing traces to generate new execution graphs for both parallelism configurations and model architecture variations, to provide accurate performance estimates through simulation. This predictive capability significantly reduces the need for costly hardware resources, positioning \tool as a practical tool for efficient model configurations exploration.

\section{Discussion}
\label{sec:discussion}

In this section, we discuss the profiling overhead, adaptability, scope, and limitations of \tool.

\textbf{Profiling Overhead and Cost.} \tool leverages PyTorch Kineto~\cite{pytorch-kineto} to collect traces and construct the execution graph. Profiling requires only a few lines of hook code, which makes it much more user-friendly than existing approaches that require additional instrumentation. Although profiling can impact execution, capturing a single iteration—or just a few—is sufficient, as the model's execution pattern remains consistent across iterations. This ensures that the overall profiling overhead is negligible in the context of the entire training process.

\textbf{Adaptability of \tool.} \tool requires profiling traces that capture both CPU and GPU activities, including framework operators, CUDA runtime events, and GPU kernels. Similar profiling capabilities are available in other ML frameworks, such as TensorFlow Profiler~\cite{TensorFlow-Profiler}. \tool's post-processing stages for constructing and manipulating execution graphs are framework-agnostic, making it easy to adapt to other frameworks.

Our methodology also extends well to other LLMs and ML models in other domains, as it does not rely on model-specific information to construct execution graphs for performance modeling. While we make certain assumptions during graph manipulation, such as where to insert layers and which tasks would be affected, our method remains broadly applicable, given the shared transformer-based architecture of most modern LLMs.

Similarly, although this paper focuses on LLM training, where communication overhead is higher and model behavior more complex, \tool is also applicable to the inference. We anticipate even broader use cases as LLM inference grows more complicated, such as distributed inference~\cite{wei2022gpu, li2023accelerating} and SSD-based inference~\cite{wilkening2021recssd, sun2022rm}.

\textbf{Kernel Execution Time Prediction.} Changing configurations can introduce new GPU kernels not present in the original trace, such as new communication kernels when adjusting parallelism strategies or new computation kernels when modifying model architectures. Accurately predicting performance for new configurations requires estimating the runtime of these new or altered kernels. Currently, we estimate the runtime of unseen kernels using an in-house GPU kernel performance model, built by analyzing fleet GPU traces, as it is available and accurate. However, alternative methods are also available. For communication kernels, we can use metadata like message size, collective algorithm, and networking environment to estimate performance with network simulators like ASTRA-sim~\cite{won2023astra} and HeterSim~\cite{tang2024simulating}, or analytical models~\cite{moolchandani2023amped, rashidi2022themis}. For computation kernels, we can simply measure runtime through individual microbenchmarks. However, predicting kernel runtimes is beyond the scope of this work.

The primary goal of \tool is to deliver a fine-grained execution graph and an accurate performance model to capture the complexities of LLM execution and provide reliable performance estimates. Developers can implement optimized individual kernels, profile their runtime, and integrate the results into \tool to predict the overall runtime, saving the engineering efforts of porting them into the frameworks. More importantly, it can offer invaluable insights for optimization even before implementation by answering what-if questions, such as how much the overall runtime would be reduced if a kernel ran twice as fast, and identifying which optimization would yield the greatest performance improvement. By modifying existing traces and estimating performance through simulation, \tool makes performance evaluation and optimization more efficient and cost-effective.

\textbf{Limitations.} \tool currently focuses on modeling and simulating the timing of model execution. In predicting performance for modified configurations, we assume the model will function as expected under the new settings, without unforeseen issues such as out-of-memory errors. Estimating system-level metrics, such as FLOPS utilization, memory consumption, bandwidth usage, or energy efficiency, lies beyond \tool's current scope. These metrics, essential for optimizing LLM efficiency, are part of our future plans to provide more comprehensive performance insights.

\section{Conclusion}

In this paper, we introduced \tool, a trace-driven performance modeling and estimation toolkit for large-scale LLM training. \tool captures complex behaviors through detailed execution graphs built from profiling traces, enabling accurate performance modeling and estimation. Our evaluation on a production ML cluster with up to 512 NVIDIA H100 GPUs shows an average replay error of only 3.3\% across diverse model architectures and parallelism configurations. By manipulating existing graphs to generate new ones for different configurations and predicting performance through simulation, \tool supports efficient optimization exploration.


\section*{Acknowledgements}
We sincerely thank the anonymous reviewers for their suggestions on earlier versions of this manuscript. This work was in part sponsored by Meta through a student researcher appointment. This work was also in part supported by NSF CAREER Award CCF-2326182, an Intel Research Award, a
Sloan Research Fellowship, a Microsoft Research Fellowship, and a Facebook Research Faculty Award.




\bibliography{ref}
\bibliographystyle{mlsys2025}



\end{document}